\documentclass[letterpaper, 10 pt, conference]{ieeeconf}  

\IEEEoverridecommandlockouts                              

\usepackage{cite}
\usepackage{amsmath,amssymb,amsfonts}
\usepackage{algpseudocode}
\usepackage{graphicx}
\usepackage{textcomp}
\usepackage{xcolor}

\usepackage[utf8]{inputenc}
\usepackage{csquotes}
\usepackage{amsmath}
\usepackage{mathtools}
\let\labelindent\relax 
\usepackage[inline]{enumitem}
\usepackage{amsfonts}
\usepackage{amssymb}

\usepackage{amsthm}
\usepackage{etoolbox}
\usepackage{siunitx}
\usepackage[hyphens]{url}
\usepackage[hidelinks, bookmarks=false]{hyperref}

\title{\LARGE \bf Characterization of the Frequency-Current Curve of First-Order High-Pass Neurons}

\renewcommand{\authorrefmark}[1]{\raisebox{0pt}[0pt][0pt]{\textsuperscript{\footnotesize #1}}}
\newcommand{\authorrefmarktanks}[1]{\raisebox{0pt}[0pt][0pt]{\textsuperscript{#1}}}

\author{Christian Fernandez Lorden\authorrefmark{1}\and Jérôme Adriaens\authorrefmark{1}\and Alessio Franci\authorrefmark{1,2}\and Gustave Bainier\authorrefmark{1}%
\thanks{\authorrefmarktanks{1}C.\ Fernandez Lorden, J.\ Adriaens, G.\ Bainier, and A.\ Franci are with the Department of Electrical Engineering and Computer Science, University of Liège, 4000 Liège, Belgium.
\authorrefmarktanks{2}A.\ Franci is with the WEL-T Department, WEL Research Institute, 1300 Wavre, Belgium. \{christian.fernandezlorden, jadriaens, afranci, gustave.bainier\}@uliege.be.}
}
\date{November 2026}

\setlist[enumerate]{label=\arabic*), ref=\arabic*)}

\newcommand{\R}{\mathbb R}
\newcommand{\dx}[1]{\mathop{\mathrm{d}#1}}
\newcommand{\eqdefsymbol}{\coloneqq}

\newtheorem{theorem}{Theorem}
\newtheorem{lemma}[theorem]{Lemma}
\newtheorem{corollary}[theorem]{Corollary}
\newtheorem{proposition}[theorem]{Proposition}

\newtheorem{definition}[theorem]{Definition}
\newtheorem{example}[theorem]{Example}

\newif\ifwithappendix
\withappendixtrue

\makeatletter
\g@addto@macro\@floatboxreset{\vspace*{5pt}}
\patchcmd{\table}{\@setminipage}{\@setminipage\vspace*{5pt}}{}{\PackageWarning{main}{table patch failed}}
\expandafter\patchcmd\csname table*\endcsname{\@setminipage}{\@setminipage\vspace*{5pt}}{}{\PackageWarning{main}{table* patch failed}}
\patchcmd{\@maketitle}{\vskip0.25in}{\vskip0.25in\vskip3pt}{}{\PackageWarning{main}{title patch failed}}
\makeatother

\begin{document}
\bstctlcite{BSTcontrol} 

\maketitle

\begin{abstract}
    A spiking model, called the High-Pass (HP) model, has recently been proposed as a minimal excitable primitive.
    This model is composed of a static activation function, usually a sigmoid, in positive feedback with a high-pass filter.
    The HP model enters a tonic spiking regime whenever its input lies in the high-gain zone of the sigmoid, where the sigmoid's slope exceeds 1.
    In neuroscience, the frequency-current (FI) curve is the standard way of characterizing the tonic spiking regime of a neuron.
    So far, the FI curve of the HP model has been characterized analytically only for specific sigmoids, and has otherwise been computed through simulation.
    These simulations are computationally heavy, since simulating neurons requires solving a stiff system of differential equations.
    This paper bridges this gap by analytically characterizing the FI curve of the HP model with a first-order high-pass filter, for a broad class of sigmoid functions.
    The FI curve of the first-order HP model can be computed by solving two input-independent equations and evaluating two input-dependent integrals, with no need to simulate the model itself.
    A spiking primitive is said to be Type~I (resp. Type~II) if tonic spiking starts at a zero (resp. non-zero) frequency.
    This paper also shows that, for the HP model, the type depends solely on the local shape of the sigmoid at the outer edges of its high-gain zone.
    A sharp corner, where the derivative approaches a value strictly smaller than 1 from the outside, yields a Type~I primitive. A locally twice differentiable edge with a bounded second derivative and a derivative approaching 1, yields a Type II primitive.
\end{abstract}

\section{Introduction}\label{sec:introduction}

The frequency-current (FI) curve, the steady-state firing rate of a neuron as a function of a constant input, is the standard characterization of tonic spiking.
It goes back to Hodgkin's study of crab axons~\cite{hodgkin_PIcurve_1948}, where it appears in a reciprocal form: a period-current curve.
Hodgkin studied repetitive firing and sorted axons into classes based on the width of their firing frequency range: class~I axons fire over a wide range, from low to high frequencies, whereas class~II axons only fire in a narrow high-frequency band.
Over time, these classes have been idealized as follows: Type~I neurons start firing at an arbitrarily low frequency, whereas Type~II neurons start at a non-zero frequency.
Dynamical systems theory later linked these types to the bifurcation at the static onset of firing~\cite{rinzelermentrout_neuralexcitation_1989, izhikevich_dynamical_2006}, with Type~II arising from an Andronov-Hopf bifurcation and Type~I from either a saddle-node on an invariant circle or a saddle homoclinic orbit bifurcation.
Traditionally, the FI curve of most neuron models is obtained either analytically, for specific parameters or activation functions~\cite{tonnelier_piecewisemorris_2003}, or numerically, by simulating the model at each constant input~\cite{worm_inward_2026} or by numerical continuation of the limit cycle~\cite{rush_potassium_1995}.
The first method is case-specific and thus labor-intensive, while the second requires a large amount of computation to sample the FI curve and may become numerically unstable for very slow firing.

\begin{figure}[t]
    \centering
    \includegraphics[width=0.8\columnwidth]{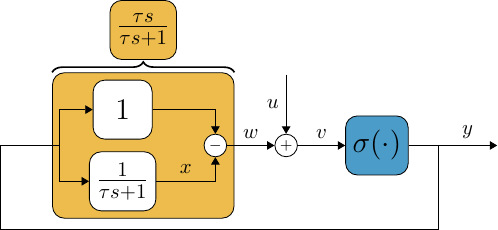}
    \caption{Block diagram of a High-Pass (HP) neuron. Adapted from \cite{Bainier2026_poster}.}
    \label{fig:hp-neuron}
\end{figure}

The neuron model we consider in this paper is a new spiking model called the \emph{High-Pass (HP) model}, first presented in 2026 in~\cite{Bainier2026_poster}.
It is composed of a sigmoid function in positive feedback with a high-pass filter (see Fig.~\ref{fig:hp-neuron} for the first-order model).
Like the FitzHugh-Nagumo, Morris-Lecar~\cite{izhikevich_models_2004}, or mixed-feedback models~\cite{ribar_neuromorphiccontrol_2021}, this model is a simplified continuous model of neural excitability. It may in fact be viewed as a practical rewriting of a FitzHugh-Nagumo model~\cite{Bainier2026_poster,BAINIER_ECC}.
Unlike the other models, the HP model decouples the nonlinearity from the dynamics: the dynamics are completely linear, and the excitable behavior is obtained with a single static nonlinearity.
This allows for a frequency-based understanding of excitability: spikes come from local high gain combined with a feedback amplification of the high frequencies of the output.
The model also extends to bursting~\cite{ARTHUR_ECC} and can be implemented in electronics~\cite{Bainier2026_poster}, as sigmoids and linear filters can be realized using conventional electronic circuits.

Given the simplicity of the HP model, this paper characterizes its tonic spiking regime for a first-order high-pass filter and a large class of sigmoids, including non-smooth ones. We show that, for any suitable sigmoid, the first-order HP neuron enters tonic spiking whenever the constant input lies in the \emph{high-gain zone} of the sigmoid, and converges to a stable resting state when the input is outside that zone. Our main contribution is an implicit formula that can be used to numerically or algebraically compute the FI curve of any first-order HP neuron. The formula is tractable, and can be evaluated by solving two input-independent equations and evaluating two input-dependent integrals, with no need to simulate the model itself. Moreover, the formula yields sufficient conditions on the sigmoid to determine whether the model is Type~I or Type~II, based only on its local shape near the outer edges of the high-gain zone.

The paper is structured as follows. Section~\ref{sec:preliminaries} introduces the notation and our definition of a sigmoid. Section~\ref{sec:motif} introduces the HP model and a change of variables that simplifies its analysis. Section~\ref{sec:ifcurve} studies the limit cycle and provides a tractable formula for the FI curve, together with two examples. Section~\ref{sec:typeI} provides and illustrates sufficient conditions for a Type~I or Type~II model. Section~\ref{sec:conclusion} concludes the paper and presents avenues for further research.\ifwithappendix\ Appendix~\ref{appendix:proofs} contains the proofs of all results.\fi

Code and data are on GitHub: \url{https://github.com/ChristianFernandezLorden/ECC27-FernandezLorden-HighPassNeuronFICurve}.

\section{Preliminaries}\label{sec:preliminaries}
We denote by $\R$ the set of real numbers and by $C^n$ the set of functions from $\R$ to $\R$ that are $n$ times continuously differentiable.
\begin{definition}[Sigmoid]\label{def:sigmoid}
    A sigmoid function, denoted $\sigma(x)$, is a function that satisfies the following properties:
    \begin{enumerate}
        \item \emph{Continuity:} $\sigma\in C^0$\textup{;}\label{item-def:sigmoid-cont}
        \item \emph{Single high-gain zone:} There exists a single pair $a<b$ such that $\sigma(x)-x$ is strictly decreasing on $(-\infty,a)$ and on $(b,+\infty)$, and strictly increasing on $(a,b)$\textup{;}\label{item-def:sigmoid-gain-zone}
        \item \emph{Linear-dominated:} $\lim_{x\rightarrow\pm\infty}(\sigma(x)-x)=\mp\infty$\textup{.}\label{item-def:sigmoid-linear-dom}
    \end{enumerate}
    We denote by $\mathcal{S}$ the class of all sigmoid functions and refer to the interval $(a,b)$ as the high-gain zone of the sigmoid.
\end{definition}

The previous definition implies the following properties.
\begin{corollary}[Consequences of the sigmoid definition]\label{corollary:sigmoid}
    Any sigmoid $\sigma\in\mathcal{S}$ satisfies the following properties:
    \begin{enumerate}
        \item The derivative $\sigma'$ exists almost everywhere on $(-\infty,\infty)$;\label{item-cor:sigma-deriv}
        \item In the high-gain zone, $\sigma$ is strictly increasing, with $\sigma'(x)\ge1$ where $\sigma'$ exists.\label{item-cor:sigma-hgzmonotone}
    \end{enumerate}
\end{corollary}

\section{The high-pass motif}\label{sec:motif}

We consider the \emph{High-Pass (HP) neuron}~\cite{Bainier2026_poster} illustrated in Fig.~\ref{fig:hp-neuron}. In state-space form, it reads:
\begin{equation}
    \tau \dot{x}=-x+\sigma(v), \qquad v = u + \sigma(v) - x, \label{eq:high-pass}
\end{equation}
with $\sigma \in \mathcal{S}$ and $\tau\in\mathbb{R}_{>0}$, where $x(t)\in\R$ is the state of the low-pass filter from which the high-pass filter is built, $v(t)\in\R$ is the input of the sigmoid, and $u(t)\in\R$ is the input of the neuron. The output of the neuron is defined as $y(t)\eqdefsymbol\sigma(v(t))$. Since our goal is to analyze the FI curve of this model, which is defined using constant inputs, we assume a constant input $u(t) \equiv u$. This allows for an advantageous reparameterization of the neuron.
Setting $z\eqdefsymbol x-u$ yields:
\begin{equation}
    \tau \dot{z}=v-u, \qquad v = \sigma(v) - z.\label{eq:high-pass-repar}
\end{equation}
This reparameterization makes the study of the limit cycle easier in the phase plane because, as Fig.~\ref{fig:phase-plane} shows, the $z$ nullcline becomes a vertical line in the $(v,z)$ plane.

The HP neuron is acausal. However, it can be seen as the singular limit of a two-dimensional causal dynamics obtained by applying an arbitrarily fast low-pass dynamics to $v$:
\begin{equation}
    \tau \dot{z}=v-u, \qquad \varepsilon\dot v = \sigma(v) - v - z,\label{eq:high-pass-repar-regul}
\end{equation}
with $\varepsilon\ll\tau$.
The practical behavior of the original acausal system \eqref{eq:high-pass-repar} is recovered in the singular limit $\varepsilon\to0$, when the $v$ dynamics are instantaneous compared with the $z$ dynamics.

The function $\sigma(v) - v$, which is the $v$ nullcline of \eqref{eq:high-pass-repar-regul}, appears regularly in the study of the oscillation period.
We therefore assign it a symbol, $\psi(v)\eqdefsymbol \sigma(v) - v$, and call it the S-curve of the model.

\section{The implicit FI curve}\label{sec:ifcurve}

\begin{figure}
    \centering
    \includegraphics[width=0.9\columnwidth]{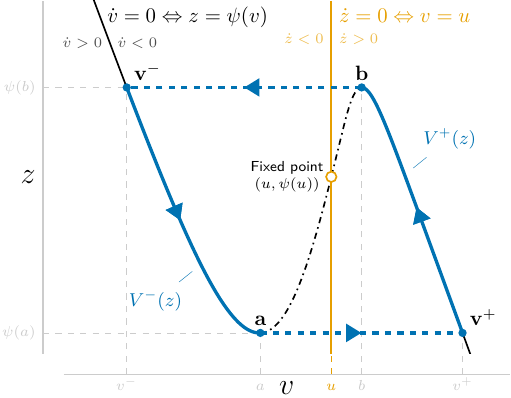}
    \caption{Typical phase plane of an HP neuron in the $(v,z)$ plane. The nullclines of $z$ and $v$ are shown in black and orange, respectively, and the spiking limit cycle in blue.}
    \label{fig:phase-plane}
\end{figure}

A natural way of studying the tonic spiking oscillations of the HP neuron is to look at its phase plane, as illustrated in Fig.~\ref{fig:phase-plane}. The oscillation cycles between $(a,\psi(a))$, $(v^+,\psi(a))$, $(b,\psi(b))$, and $(v^-,\psi(b))$. The points $a$ and $b$ correspond to the abscissas of the local minimum and maximum of the S-curve, while $v^-$ and $v^+$ denote the abscissas of the points obtained by horizontally projecting these extrema onto the exterior branches of the S-curve. This figure holds qualitatively for any sigmoid satisfying Definition~\ref{def:sigmoid}, and the following lemma guarantees the existence of $v^-$ and $v^+$.

\begin{lemma}[Spike terminating points]\label{lemma:cycle_points_and_inverses}
    Consider the HP neuron system \eqref{eq:high-pass-repar} and any sigmoid $\sigma\in\mathcal{S}$.
    There exist unique points $v^-$ and $v^+$ that satisfy $v^-<a$, $v^+ > b$, and
    \begin{equation}
        \psi(v^-)=\psi(b),\qquad\psi(v^+)=\psi(a).\label{eq:v-variables-def}
    \end{equation}
    There also exist unique inverse functions, $ V^-:[\psi(a),\infty)\to(-\infty,a]$ and $V^+:(-\infty,\psi(b)]\to[b,\infty)$, of the left and right outer branches of the $v$ nullcline $\psi(v)$ (see Fig.~\ref{fig:phase-plane}).
    They satisfy 
    \begin{equation}
       V^-(\psi(v)) = v, \qquad V^+(\psi(v)) = v,\label{eq:v-function-inverse}
    \end{equation}
    for $v\in(-\infty,a]$ and $v\in[b,\infty)$ respectively.
\end{lemma}


Since we consider the dynamics of $v$ to be instantaneous ($\varepsilon\to0$), the system only lives on the $v$ nullcline, and the oscillation period is fully determined by the time spent on the left and right branches $V^-$ and $V^+$. If we can determine the oscillation period $T(u)$ then we easily get the FI curve $f(u)$ as $f(u)=1/T(u)$. 
Fortunately, the derivative of $z$ along those branches is well defined.
It can be expressed as:
\begin{equation}
    \dot z = \frac{V^\mp(z)-u}{\tau}, \label{eq:branch-z-deriv}
\end{equation}
on the left and right branches respectively. Using a method similar to \cite[Section V-A]{sacre_singularly_2016}, the following theorem characterizes the period of the neuron as a function of the input. 

\begin{theorem}[Spiking period] \label{theorem:spiking-period}
    Consider the HP neuron system \eqref{eq:high-pass-repar} and any sigmoid $\sigma\in\mathcal{S}$.
    \begin{enumerate}
        \item For $u\in(-\infty,a)\cup(b,\infty)$, there is no limit cycle: there exists a unique stable resting point $(v,x) = (u,\sigma(u))$.\label{item-thm:stability-spiking-period}
        \item For $u\in(a,b)$, there exists a unique stable limit cycle with period\label{item-thm:cycle-spiking-period}
        \begin{equation}
            \begin{aligned}
            \!\!T(u) &= T^-(u)+T^+(u)\\ &= \tau \left(\int_{\psi(b)}^{\psi(a)}\frac{\dx{z}}{V^-(z)-u} + \int_{\psi(a)}^{\psi(b)}\frac{\dx{z}}{V^+(z)-u} \right).\label{eq:period-int}
            \end{aligned}
        \end{equation}
    \end{enumerate}
\end{theorem}

\begin{figure}
    \centering
    \includegraphics[width=0.9\columnwidth]{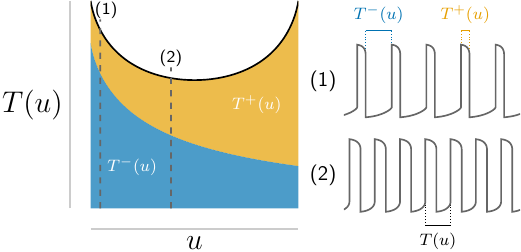}
    \caption{Limit cycle period $T(u)$ of an HP neuron with $\sigma(v)=\tanh(2v)$. The times spent on the left branch, $T^-(u)$, and on the right branch, $T^+(u)$, are shown as blue and orange areas, respectively. The traces on the right show the neuron output $\sigma(v)$ for two values of $u$.}
    \label{fig:period}
\end{figure}

Fig.~\ref{fig:period} illustrates the period $T(u)$ and its decomposition into $T^-(u)$ and $T^+(u)$.
One thing that stands out from this theorem is that the timescale $\tau$ of the high-pass filter multiplies the entire period.
This means that the parameterization of the first-order high-pass filter only rescales the FI curve, while its shape is determined by the sigmoid alone.

The characterization of the spiking period in Theorem~\ref{theorem:spiking-period} is informative about the neuron behavior, but it is a poor way of computing the spiking period for an arbitrary sigmoid.
Indeed, it requires evaluating $V^-$ and $V^+$, the local inverses of the S-curve $\psi$, which cannot always be found analytically, and generally require a computationally expensive root-finding evaluation. The following proposition shows that the same period can be computed using solely the S-curve $\psi$ and the terminating points of the spikes $v^-$ and $v^+$.

\begin{proposition}[Tractable spiking period] \label{proposition:spiking-period-compute}
    Consider the HP neuron system \eqref{eq:high-pass-repar} and any sigmoid $\sigma\in\mathcal{S}$.
    Let $u\in(a,b)$.
    Then the period of the stable limit cycle can be computed as
    \begin{equation}
    \begin{aligned}
        T(u) =\;& \tau(\psi(b)-\psi(a))\left(\frac{1}{v^+-u}+\frac{1}{u-v^-}\right)\\
        &+ \tau\int_{v^-}^{a}\frac{\psi(v)-\psi(a)}{(u-v)^2}\dx{v}\\
        &+ \tau\int_{b}^{v^+}\frac{\psi(b)-\psi(v)}{(v-u)^2}\dx{v}.
    \end{aligned}\label{eq:period-int-tractable}
    \end{equation}
\end{proposition}

\begin{figure}
    \centering
    \includegraphics[width=0.9\columnwidth]{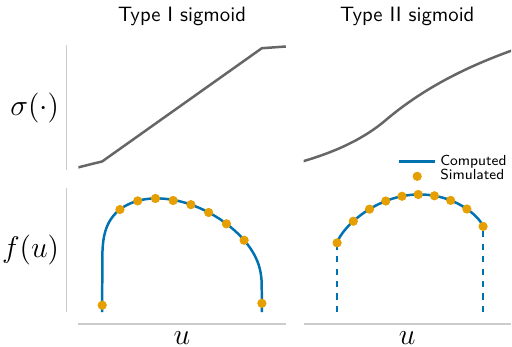}
    \caption{Sigmoids (top) and FI curves (bottom) of a Type~I and a Type~II HP neuron. The FI curves are computed with formula \eqref{eq:period-int-tractable} and validated in simulation with a regularization timescale $\varepsilon=10^{-8}\tau$. Dashed lines show the non-zero onset frequencies of the Type~II neuron. The Type~I sigmoid is the same as Example \ref{example:lin_part} with $k_l=0.7,k_m=2,k_r=0.2$ and the Type~II sigmoid is the same as Example \ref{example:smooth_sig} with $k=2,m_l=0.6,m_r=1.2$.}
    \label{fig:type}
\end{figure}

{Table}~\ref{tab:compute_time} compiles the average compute time of our Julia code for FI curves of Fig.~\ref{fig:type}, computed using both formula~\eqref{eq:period-int-tractable} and direct simulations of the model. The table highlights that our formula is around 2 times faster than doing simulations, while evaluating 100 times more points. Computing a point is therefore more than two orders of magnitude cheaper using the formula as compared to running simulations.

\begin{table}
    \centering
    \renewcommand{\arraystretch}{1.2}
    \caption{Computation Time of the FI Curves of Fig. \ref{fig:type} across 100 Trials}
    \begin{tabular}{l|cc}
         Compute Time (\unit{\milli\second}, mean $\pm$ std) & Type~I               & Type~II\\\hline
         Formula~\eqref{eq:period-int-tractable} (1000 points)  & \num{89.3(8.6)}  & \num{75.1(5.8)}\\
         Simulations (10 points) & \num{156.4(25.4)} & \num{191.4(31.5)}
    \end{tabular}
    \label{tab:compute_time}
\end{table}

The structure of the tractable formula~\eqref{eq:period-int-tractable} makes very clear that the spiking period depends only on the shape of the sigmoid outside the high-gain zone. Moreover, it also allows us to show that the FI curve of the HP model is smooth and unimodal: it starts at some value at $a$, smoothly rises to a maximal frequency, then decreases toward some value at $b$ (the maximum may also lie at an edge). Since the non-integral term of \eqref{eq:period-int-tractable} is always strictly positive, the spiking frequency can never become arbitrarily high.

\begin{corollary}[Properties of the FI curve] \label{corollary:FIcurve-cont}
     For any sigmoid $\sigma\in\mathcal{S}$, the FI curve of the HP neuron system \eqref{eq:high-pass-repar} is smooth and strictly positive on $u\in(a,b)$ and admits at most a single local maximum.
\end{corollary}

Since both \eqref{eq:period-int} and \eqref{eq:period-int-tractable} are analytical formulations, they may also be used to compute the explicit FI curve for specific sigmoids. Below, we consider two sigmoids for which the explicit FI curve can be obtained. 

\begin{example}[Piecewise-linear sigmoid]\label{example:lin_part}
    Consider the sigmoid $\sigma$ defined in the following way
    \begin{equation*}
        \sigma(v) = \begin{cases}
            k_l v + k_l - k_m, & v\in (-\infty ,-1)\\
            k_m v, & v\in[-1,1]\\
            k_r v  - k_r + k_m, & v\in (1,\infty)
        \end{cases}
    \end{equation*}
    with $k_l,k_r\in[0,1)$ and $k_m\in(1,\infty)$. 
    It satisfies the condition of Definition \ref{def:sigmoid} and we can derive that $a=-1$, $b=1$, $v^- = -1-{2(k_m-1)}/{(1-k_l)}$, and $v^+=1+{2(k_m-1)}/{(1-k_r)}$. 
    Using \eqref{eq:period-int} and standard techniques to evaluate the integrals, one can determine its analytic period curve as
    \begin{equation*}
        \begin{aligned}
        T(u) = &\tau(1-k_l)\ln\left(1+\frac{2(k_m-1)}{(1-k_l)(1+u)}\right) \\&+ \tau(1-k_r)\ln\left(1+\frac{2(k_m-1)}{(1-k_r)(1-u)}\right),
        \end{aligned}
    \end{equation*}
    with $u\in(a,b)$.
\end{example}

\begin{example}[Piecewise-rational sigmoid]\label{example:smooth_sig}
    
    Consider the sigmoid $\sigma$ defined in the following way
    \begin{equation*}
        \sigma(v) = \begin{cases}
            \dfrac{kv}{1-kv/m_l}, & v\in(-\infty,0),\\[1ex]
            \dfrac{kv}{1+kv/m_r}, & v\in[0,\infty),
        \end{cases}
    \end{equation*}
    with $k\in(1,\infty)$ and $m_l,m_r\in(0,\infty)$, so that $\sigma$ saturates at $-m_l$ and $m_r$.
    It satisfies the conditions of Definition~\ref{def:sigmoid} and we can derive that $a=-m_l(\sqrt{k}-1)/k$, $b=m_r(\sqrt{k}-1)/k$, $v^-=-m_l(W_l-1)/k$, and $v^+=m_r(W_r-1)/k$, where
    \begin{equation*}
    \begin{aligned}
        W_l&=\tfrac12\big(S_l+\sqrt{S_l^2-4k}\big), &S_l&=k+1+\frac{m_r}{m_l}\big(\sqrt{k}-1\big)^2,\\
        W_r&=\tfrac12\big(S_r+\sqrt{S_r^2-4k}\big), &S_r&=k+1+\frac{m_l}{m_r}\big(\sqrt{k}-1\big)^2.
    \end{aligned}
    \end{equation*}
    Using \eqref{eq:period-int-tractable} and standard techniques to evaluate the integrals (partial fractions in $s=1+k|v|/m_l$ for $T^-(u)$ and $s=1+k|v|/m_r$ for $T^+(u)$), one can determine its analytic period curve as,
    \begin{equation*}
        T(u)=\tau\left[Q\left(W_l,\,1-\frac{ku}{m_l}\right)+Q\left(W_r,\,1+\frac{ku}{m_r}\right)\right],
    \end{equation*}
    with $u\in[a,b]$ and
    \begin{equation*}
    \begin{aligned}
        Q(W,w)=&\left(1-\frac{k}{w^2}\right)\ln(W-w)+\frac{k}{w^2}\ln\frac{W}{\sqrt{k}}\\
        &+\frac{k-w^2}{w^2}\,\ln\big(\sqrt{k}-w\big)+\frac{k}{w}\left(\frac{1}{\sqrt{k}}-\frac{1}{W}\right).
    \end{aligned}
    \end{equation*}
    The singular points $w\in \{0,\sqrt{k}\}$ are covered by continuity.
\end{example}
The proofs of the examples are omitted due to their length.

\section{Tonic spiking types}\label{sec:typeI}

The two main types of spiking cells discussed by neuroscientists are the so-called Type~I and Type~II, whose FI curves respectively start at a frequency of 0 and at a non-zero frequency~\cite{izhikevich_dynamical_2006}. We use the implicit tractable formula from the previous section to determine the type of first-order HP models based on their sigmoid.

\subsection{Types based on FI curve}

In our framework, a Type~I onset at the left edge corresponds to $\lim_{u\rightarrow a^+}T(u)=+\infty$ (resp. $\lim_{u\rightarrow b^-}T(u)=+\infty$ at the right edge).
We can derive two simple sufficient conditions, one for each type, that cover the sigmoid shapes likely to be used in practice: smooth sigmoids and sigmoids with corners.

\begin{proposition}[Type conditions]\label{prop:types-conditions}
    Consider the HP neuron system \eqref{eq:high-pass-repar} and any sigmoid $\sigma\in\mathcal{S}$.
    The onset at the left edge $a$ (resp. right edge $b$) is
    \begin{enumerate}
        \item Type~I if $\lim_{x\rightarrow a^-}\sigma'(x)=k < 1$ (resp. $\lim_{x\rightarrow b^+}\sigma'(x) = k <1$).\label{item-prop:types_conditions-typeI}
        \item Type~II if there exists $\eta> 0$ such that $\sigma\in C^2$ everywhere in a lower neighborhood $(a-\eta,a)$ (resp. an upper neighborhood $(b,b+\eta)$), the second derivative $\sigma''$ is bounded in that neighborhood, and $\lim_{x\rightarrow a^-}\sigma'(x)=1$ (resp. $\lim_{x\rightarrow b^+}\sigma'(x)=1$).\label{item-prop:types_conditions-typeII}
    \end{enumerate}
\end{proposition}

What is striking in the result above is that only the immediate outer neighborhood of the sigmoid at onset determines its type.
The rest of the sigmoid matters for the overall shape of the FI curve, but whether the model is Type~I or Type~II depends only on a local property.
The two conditions are also complementary for sigmoids that are twice differentiable with a bounded second derivative near the edge.
In that case, $\sigma'$ is Lipschitz on $(a-\eta,a)$, so its left limit at $a$ exists, and this limit is at most 1 since $\psi$ is decreasing on the left of $a$: the onset is Type~I if the limit is strictly smaller than 1, and Type~II if it equals 1.
The way we obtain a Type~I onset is similar to how \cite{tonnelier_piecewisemorris_2003} obtained theirs for a Morris-Lecar model.

Examples of Type~I and Type~II sigmoids are illustrated in Fig.~\ref{fig:type}.
The figure clearly shows that a sharp corner in the sigmoid yields a zero onset frequency, while a smooth corner yields a non-zero one.
We chose to illustrate a sigmoid with two Type~I onsets and another with two Type~II onsets, but switching from one shape to the other inside the high-gain zone would create a sigmoid with a Type~I onset on the left and a Type~II onset on the right.
Notably, for a small regularization timescale, the predicted frequency matches the observed one almost exactly.

Near the Type~I edges, numerical instabilities in the simulation of the regularized model made it hard to reach the zero frequency predicted by the formula.
This is expected, because the period grows logarithmically as the input approaches a Type~I edge, leading to an arbitrarily large sensitivity of the spiking frequency.
Indeed, for a piecewise-linear sigmoid with slope $k_l<1$, such as our Type~I example, $\psi(v)-\psi(a)=(1-k_l)(a-v)$ on $(v^-, a)$, and since
\begin{equation}
    \int_{v^-}^{a}\frac{(a-v)\dx{v}}{(u-v)^2}=\ln\frac{u-v^-}{u-a}-\frac{a-v^-}{u-v^-},
\end{equation}
formula \eqref{eq:period-int-tractable} gives $T(u)=\tau(1-k_l)\ln\frac{1}{u-a}+O(1)$ as $u\to a^+$.
As \cite{tonnelier_piecewisemorris_2003} highlights, this logarithmic onset is similar to the onset of leaky integrate-and-fire (LIF) models \cite{tal_liflog_1997} and of models with a saddle homoclinic orbit bifurcation \cite[Figure 6.30]{izhikevich_dynamical_2006}.

    

\subsection{Types based on bifurcation}

We call \emph{degenerate Type~I} the Type~I onset that this model achieves when \eqref{eq:period-int-tractable} diverges to infinity at the edge.
Indeed, the FI curve starting at 0 for a suitable sigmoid is not due to a classical Type~I bifurcation: neither a saddle-node on an invariant circle (SNIC) bifurcation nor a saddle homoclinic orbit bifurcation~\cite{izhikevich_dynamical_2006} is involved.
Those two bifurcations require a saddle coexisting with the resting point, which is impossible in a first-order high-pass model~\cite{ARTHUR_ECC}.
Instead, Type~I emerges from a Type~II bifurcation, an Andronov-Hopf bifurcation~\cite{izhikevich_dynamical_2006}, whose onset frequency degenerates to zero due to a border collision~\cite[Definition 3.4]{dibernardo_piecewisesmooth_2008} when the activation becomes non-smooth.
The degenerate Type~I differs from a classical Type~I in two major ways: no saddle-node bifurcation or saddle is involved (the equilibrium is unique for every input), and any smoothing of the sigmoid restores a non-zero onset frequency.


\section{Conclusions and perspectives}\label{sec:conclusion}

In this paper, we have studied the limit cycles of first-order HP neurons.
We have shown that tonic spiking occurs whenever the input lies in the high-gain zone of the sigmoid, and we have derived a tractable formula to compute the limit-cycle period for any input inside that zone.
We have also determined sufficient conditions to predict whether a given sigmoid gives rise to a Type~I or Type~II model, which depend only on the local properties of the sigmoid on the outer side of the onset point.

The current contribution focuses on the tonic, or steady-state, regime of first-order models.
Works such as~\cite{BAINIER_ECC,JEROME_ECC} show that the response to frequency-rich inputs is more complex than what the FI curve can capture.
Our analysis also showed that the Type~I onset obtained through a careful choice of sigmoid disappears as soon as any smoothing is applied, because the zero frequency results from a border collision rather than from one of the classical Type~I bifurcations.
This points to several interesting avenues of research.
A first would be to study the neuron and its FI curve in detail when using second- or higher-order filters.
In fact, \cite{ARTHUR_ECC} already develops part of this theory for the second-order model.
A second would be to find a minimal modification of the HP model that allows the classical Type~I bifurcations to occur, without relying on non-smooth sigmoids and border collisions.
A third would be to unify the dynamic threshold of~\cite{BAINIER_ECC} with the FI curve analysis done here, to get a full picture of the effect of the sigmoid choice on the neuron behavior.
A final, and perhaps most ambitious, direction is to develop a neuromorphic control theory built around tunable HP neurons as elementary computational primitives.

\ifwithappendix\appendix\section{Appendix} 

\subsection{Proofs}\label{appendix:proofs}

\begin{proof}[Proof of Corollary \ref{corollary:sigmoid}]
    \ref{item-cor:sigma-deriv}
    On each of the intervals $(-\infty,a)$, $(a,b)$, and $(b,\infty)$, $\sigma(x)-x$ is monotone, so it is differentiable almost everywhere by Lebesgue's Theorem~\cite[Section 6.2]{royden_realanalysis_2010}, and so is $\sigma$.
    Since these intervals cover $\R\setminus\{a,b\}$, $\sigma'$ exists almost everywhere on $\R$.
 
    \ref{item-cor:sigma-hgzmonotone}
    On $(a,b)$, $\sigma(x)=(\sigma(x)-x)+x$ is the sum of two strictly increasing functions, so it is strictly increasing, and the derivative satisfies $\sigma'(x)-1\ge0$ when it exists.
\end{proof}
 
\begin{proof}[Proof of Lemma \ref{lemma:cycle_points_and_inverses}]
    By points~\ref{item-def:sigmoid-cont} and~\ref{item-def:sigmoid-gain-zone}, $\psi$ is continuous, strictly increasing on $[a,b]$, and strictly decreasing on $(-\infty,a]$ and on $[b,\infty)$, so $\psi(a)<\psi(b)$.
    Since $\lim_{v\to\infty}\psi(v)=-\infty$ by point~\ref{item-def:sigmoid-linear-dom}, the intermediate value theorem gives a $v^+>b$ such that $\psi(v^+)=\psi(a)$, which is unique since $\psi$ is strictly decreasing on $[b,\infty)$.
    The same argument on $(-\infty,a]$, using $\lim_{v\to-\infty}\psi(v)=+\infty$, gives the unique $v^-<a$ such that $\psi(v^-)=\psi(b)$.
    Finally, $\psi$ is continuous and strictly decreasing on $(-\infty,a]$ and on $[b,\infty)$, so, on these subdomains, it is a bijection onto $[\psi(a),\infty)$ and $(-\infty,\psi(b)]$ respectively.
    The inverses of these two restrictions are $V^-$ and $V^+$, and their values at $\psi(a)$ and $\psi(b)$ follow from those of $\psi$ at $a$, $b$, $v^-$, and $v^+$.
\end{proof}

\begin{proof}[Proof of Theorem \ref{theorem:spiking-period}]
    In the coordinates $\tilde v = v-u$ and $\tilde z = z-\psi(u)$, which shift the equilibrium $(u,\psi(u))$ of the causal model \eqref{eq:high-pass-repar-regul} to the origin, the model reads
    \begin{equation}
        \tau\dot{\tilde z}=\tilde v,\qquad \varepsilon\dot{\tilde v} = \psi(\tilde v + u) - \psi(u) - \tilde z.
    \end{equation}
    The positive definite function $V(\tilde v, \tilde z) \eqdefsymbol \big(\varepsilon {\tilde v}^2 + \tau {\tilde z}^2\big)/2$ satisfies $ \dot V = \tilde v\left(\psi(\tilde v + u) - \psi(u)\right).$
    If $\psi$ is strictly decreasing (resp. increasing) near $u$, then $\dot V\le0$ (resp. $\dot V\ge0$) near the origin, with equality only on the line $\tilde v=0$.
    On this line, $\varepsilon\dot{\tilde v}=-\tilde z$, so no solution other than the origin can stay on it.
    In the singular limit $\varepsilon\to 0$, motions in the $v$ direction become jumps, and the state slides along the outer branches of the nullcline $\psi$ following \eqref{eq:branch-z-deriv}.
 
    \ref{item-thm:stability-spiking-period}
    For $u\in(-\infty,a)\cup(b,\infty)$, $\psi$ is strictly decreasing near $u$, so $\dot V \le 0$ near the origin. 
    LaSalle's invariance principle~\cite[Corollary 4.1]{khalil_nonlinear_2002} then shows that the equilibrium is locally asymptotically stable.
    Assume $u<a$; the case $u>b$ is symmetric.
    The equilibrium then lies on the left branch.
    In the singular limit, a state starting with $z>\psi(b)$ jumps directly to the left branch and follows it to the equilibrium, and one starting with $z<\psi(a)$ jumps to the right branch, which carries it up to $(b,\psi(b))$, from where it jumps to the left branch at $(v^-,\psi(b))$.
    A state starting with $z\in[\psi(a),\psi(b)]$ jumps to one of the two branches and the same reasoning applies.
 
    \ref{item-thm:cycle-spiking-period}
    For $u\in(a,b)$, $\psi$ is strictly increasing near $u$, so $\dot V \ge 0$ near the origin. 
    If a solution starting in a small closed ball around the origin, but not at the origin, stayed in this ball, LaSalle's theorem~\cite[Theorem 4.4]{khalil_nonlinear_2002} applied with $-V$ would make it converge to the origin.
    This is impossible, since $V$ is non-decreasing along the solution, starts positive, and vanishes only at the origin, so the trajectory must leave the ball. 
    Therefore, the equilibrium is unstable.

    In the singular limit, the equilibrium lies on the middle branch, between the two outer branches, and is never reached.
    The state slides down the left branch to $(a,\psi(a))$, jumps to $(v^+,\psi(a))$, slides up the right branch to $(b,\psi(b))$, and jumps back to $(v^-,\psi(b))$.
    For a smooth S-curve with non-degenerate folds, the convergence of the limit cycle of \eqref{eq:high-pass-repar-regul} to this singular cycle as $\varepsilon\to0$ is classical~\cite[Theorem 2.1]{krupa_relaxation_2001}.
    Separating variables in \eqref{eq:branch-z-deriv} gives the time spent on each branch,
    \begin{equation}\label{eq:arc-time}
         \!\!T^-\!(u) =\!\! \int_{\psi(a)}^{\psi(b)}\!\!\!\frac{\tau\dx{z}}{u-V^-(z)},\,
         T^+\!(u) = \!\!\int_{\psi(a)}^{\psi(b)}\!\!\!\frac{\tau\dx{z}}{V^+(z)-u},
    \end{equation}
    which are positive and finite since $V^-(z)\le a<u<b\le V^+(z)$.
    Swapping the bounds of $T^-(u)$ yields \eqref{eq:period-int}.
\end{proof}

\begin{proof}[Proof of Proposition \ref{proposition:spiking-period-compute}]
    We derive the expression of $T^-(u)$; the derivation of $T^+(u)$ is analogous.
    For $z\in[\psi(a),\psi(b)]$, we have $V^-(z)\in[v^-,a]$, and since $\psi$ is strictly decreasing on $[v^-,a]$, $v<V^-(z)$ if and only if $\psi(v)>z$.
    So, with $\mathrm{H}$ the Heaviside function, we develop the integrand of $T^-(u)$ as
    \begin{equation}
        \begin{split}
        \frac{1}{u-V^-(z)}&= \frac{1}{u-v^-}+\int^{V^-(z)}_{v^-}\frac{\dx{v}}{(u-v)^2}\\&= \frac{1}{u-v^-}+\int^{a}_{v^-}\frac{\mathrm{H}(V^-(z)-v)}{(u-v)^2}\dx{v}\\&= \frac{1}{u-v^-}+\int^{a}_{v^-}\frac{\mathrm{H}(\psi(v)-z)}{(u-v)^2}\dx{v}.
        \end{split}
    \end{equation}
    We multiply by $\tau$ and integrate over $z\in[\psi(a),\psi(b)]$ to obtain $T^-(u)$ as in \eqref{eq:arc-time}.
    Since the integrand is non-negative, Tonelli's Theorem~\cite[Section 20.1, Corollary 7]{royden_realanalysis_2010} allows swapping the order of integration, and since $\int_{\psi(a)}^{\psi(b)}\mathrm{H}(\psi(v)-z)\dx{z}=\psi(v)-\psi(a)$ for $v\in[v^-,a]$, we get
    \begin{equation}
        T^-(u)=\frac{\tau(\psi(b)-\psi(a))}{u-v^-}+\tau\int_{v^-}^{a}\frac{\psi(v)-\psi(a)}{(u-v)^2}\dx{v}.
    \end{equation}
    Adding the analogous expression of $T^+(u)$ yields \eqref{eq:period-int-tractable}.
\end{proof}

\begin{proof}[Proof of Corollary \ref{corollary:FIcurve-cont}]
     For $u\in(a,b)$, every $v$ in the integration ranges of \eqref{eq:period-int-tractable} lies outside $(a,b)$, so $1/(v^+-u)$, $1/(u-v^-)$, and $1/(u-v)^2$ are positive and strictly convex in $u$, while $\psi(b)-\psi(a)>0$, $\psi(v)-\psi(a)\ge0$, and $\psi(b)-\psi(v)\ge0$.
    By dominated convergence on compact subsets of $(a,b)$, each term of \eqref{eq:period-int-tractable} is thus continuous and convex in $u$, and the first one is strictly positive and strictly convex.
    Hence $T$ is continuous, positive, and strictly convex on $(a,b)$, so $f=1/T$ is continuous, positive, and has at most one local maximum.
    The same argument applied to the $u$-derivatives of the integrands, which stay bounded on compact subsets of $(a,b)$, shows that $T$ is smooth.
\end{proof}

\begin{proof}[Proof of Proposition \ref{prop:types-conditions}]
    We treat the edge $a$; the edge $b$ is analogous.
    As $u\to a^+$, the first and third terms of \eqref{eq:period-int-tractable} stay bounded, since $v^-<a<b<v^+$.
    The integrand of the second term is non-negative and increases as $u$ decreases to $a$.
    By the Monotone Convergence Theorem~\cite[Section 4.3]{royden_realanalysis_2010}, applied along any sequence of inputs decreasing to $a$, this term therefore converges to $\tau$ times
    \begin{equation}
        \int_{v^-}^{a}\frac{\psi(v)-\psi(a)}{(a-v)^2}\dx{v}.\label{eq:integral_typeI}
    \end{equation}
    The onset is thus Type~I if \eqref{eq:integral_typeI} is infinite, and Type~II otherwise.
    Since the integrand of \eqref{eq:integral_typeI} is continuous on $[v^-,a)$, only its integral over $(a-\eta,a)$ for any $\eta\in(0,a-v^-]$ matters.
 
    \ref{item-prop:types_conditions-typeI}
    Since $\psi$ is decreasing on $[v^-,a]$, a corollary of Lebesgue's Theorem~\cite[Section 6.2, Corollary 4]{royden_realanalysis_2010} gives $\psi(v)-\psi(a)\ge\int_{v}^{a}\big(1-\sigma'(l)\big)\dx{l}$.
    Since $\sigma'(x)\to k<1$ as $x\to a^-$, there exists $\eta\in(0,a-v^-]$ such that $1-\sigma'(l)\ge(1-k)/2$ wherever $\sigma'$ is defined on $(a-\eta,a)$.
    Hence $\psi(v)-\psi(a)\ge(1-k)(a-v)/2$ on $(a-\eta,a)$, and
    \begin{equation}
        \int_{a-\eta}^{a}\frac{\psi(v)-\psi(a)}{(a-v)^2}\dx{v} \ge \frac{1-k}{2}\int_{a-\eta}^{a}\frac{\dx{v}}{a-v}=\infty.\label{eq:typeI_bound}
    \end{equation}
 
    \ref{item-prop:types_conditions-typeII}
    Reducing $\eta$ if necessary, we can assume $\eta\le a-v^-$, and we let $M\eqdefsymbol\sup_{w\in(a-\eta,a)}|\sigma''(w)|$.
    For $l\in(a-\eta,a)$, integrating $\sigma''$ over $[l,a-t]$ and letting $t\to0^+$ gives $1-\sigma'(l)\le M(a-l)$, since $\lim_{x\to a^-}\sigma'(x)=1$.
    Moreover, $1-\sigma'(l)\ge0$ since $\psi$ is decreasing on $(a-\eta,a)$.
    The Fundamental Theorem of Calculus on $[v,a-t]$, with $t\to0^+$ and the continuity of $\psi$ at $a$, then gives, for $v\in(a-\eta,a)$,
    \begin{equation}
        \psi(v)-\psi(a)=\int_{v}^{a}\big(1-\sigma'(l)\big)\dx{l} \le\frac{M}{2}(a-v)^2,
    \end{equation}
    which means $\int_{a-\eta}^{a}\frac{\psi(v)-\psi(a)}{(a-v)^2}\dx{v} \le \frac{M}{2}\eta<\infty.$
\end{proof}

\fi

\section*{Acknowledgment}
\addcontentsline{toc}{section}{Acknowledgment}

Claude Opus 5.5 (Anthropic) was used to edit and improve the phrasing of the text throughout this article, and to compute the integrals of Examples~\ref{example:lin_part} and~\ref{example:smooth_sig}.
All other technical content, results, and proofs are the authors' own.
The authors reviewed all resulting text and are responsible for the content of this article.

\bibliographystyle{IEEEtran}
\bibliography{refs}

\end{document}